# On Outage Capacity for Incremental Relaying with Imperfect Feedback

Tobias Renk, Holger Jäkel, Friedrich. K. Jondral
Communications Engineering Lab
Karlsruhe Institute of Technology (KIT)
Email: tobias.renk@kit.edu

***Abstract*—We investigate the effect of imperfect feedback on the $\epsilon$-outage capacity of incremental relaying in the low signal-to-noise ratio (SNR) regime. We show that imperfect feedback leads to a rescaling of the pre-log factor (comparable to the multiplexing gain for networks operating in the high SNR regime) and thus reduces the $\epsilon$-outage capacity considerably. Moreover, we investigate the effect of different degrees of feedback reliability on the system performance. We further derive a simple binary tree-based construction rule to analyze networks with an arbitrary number of relay nodes with respect to imperfect feedback. This rule can directly be mapped to a comprehensive matrix notation.**

## I. Introduction

The idea of incremental relaying has first been published in [1]. Incremental relaying describes a cooperative protocol where the relay only aids communication, if it has received a request from the destination. This leads to a more efficient use of the degrees of freedom of the channel. For instance, consider a relay network consisting of one source $\mathrm{S}$, one relay $\mathrm{R}$, and one destination $\mathrm{D}$. Assume that the overall transmission block is divided into two sub-blocks of equal length. In the first sub-block the source transmits its message to the destination. Due to the broadcast nature of the wireless channel, the relay is also able to receive the source message. After the first sub-block the destination tries to decode the source message. If it has been able to decode, it broadcasts a positive feedback, e.g., a one-bit feedback set to $1$, indicating successful transmission. Hence, no need by the relay is required and the second sub-block can be used by the source again in order to transmit the next message. If the destination has not been able to decode, it broadcasts a negative feedback, e.g., a one-bit feedback set to $0$. As a consequence, the second sub-block is now allocated by the relay that transmits a version of the source message depending on the cooperative strategy (amplify-and-forward, decode-and-forward, and compress-and-forward). Summarizing, with respect to outage performance, incremental relaying is at least as good as a comparable network employing the same cooperative strategy where the relay always transmits in the second sub-block.

The $\epsilon$-outage capacity of a decode-and-forward (DF) cooperative network with incremental relaying in the low signal-to-noise ratio (SNR) regime has been investigated in [2]. There, it has been shown that the protocol is outage optimal if the relay is located close to the source. Outage optimal in that case refers to the fact that the protocol achieves the $\epsilon$-outage capacity of the cut-set bound. However, if the relay is located close to the destination, there is a gap between the $\epsilon$-outage capacities of the cut-set bound and the DF protocol. In [3] the $\epsilon$-outage capacity of a bursty version of the amplify-and-forward (BAF) protocol with incremental relaying has been derived. This protocol is based on the idea that in the low power regime pulse position modulation with a very low duty cycle approaches the capacity of an ideal bandlimited additive white Gaussian noise channel (see [4], [5]). The same approach has been used by Avestimehr and Tse to derive the $\epsilon$-outage capacity of the frequency division duplex Rayleigh channel [6]. However, the authors did not consider incremental relaying.

In [2] and [3], feedback from the destination is perfectly received at the relay and the source. Hence, each node knows exactly what to do after the destination has sent the feedback and each node always does the right thing, i.e., there would never be some kind of collision due to the fact that the source and the relay access the channel simultaneously. This changes, however, if the feedback is not considered to be perfect any more. There a numerous possible scenarios if the feedback is imperfect, for instance:

- The relay remains silent through the second sub-block although it should transmit. This finally leads to an outage.
- The source retransmits its message and the relay does not, even if it has a better channel to the destination. This leads to a lower decoding probability at the destination.
- In general, the feedback link to the relay differs from the one to the source. Therefore, collisions can occur, when both terminals transmit in the second sub-block.

The questions addressed in this paper are the following: What happens if the feedback is imperfectly received at the source and the relay? Especially, how does imperfect feedback influence the $\epsilon$-outage capacity? We summarize our results: The quality of the feedback link has a strong influence on the average amount of transmission phases, which determine the pre-log factor (i.e., the scaling factor in front of the $\log$- function of capacity expressions [7]). By modeling the feedback link as a binary symmetric channel, which is reasonable since we have a one-bit feedback, we are able to quantify the pre-log factor and, thus, the $\epsilon$-outage capacity of various cooperative networks with incremental relaying.

The remainder of the paper is organized as follows. In Section II we explain the system model. Section III deals with the average amount of transmission phases per source message due to imperfect feedback. Especially, we consider the one-relay case and an extension to an arbitrary number of $K$ relays. Finally, Section IV summarizes our results and concludes the paper.

## II. System Model and Preliminaries

We consider networks consisting of one source S, $K$ relays $\{\mathrm{R}_k\}_{k=1}^{K}$, and one destination D. The receive signal at the destination after one transmission block of $T$ channel uses depends on success or failure of prior transmissions. For instance, consider the one-relay case. In the first $T/2$ channel uses the source transmits with rate $2R$ to the destination. If source transmission failed, the relay will access the channel for the remaining $T/2$ channel uses transmitting an alternate version of the source signal also with rate $2R$. The initial rate of $2R$ is due to the fact that the overall amount of transmitted information compared to direct transmission with rate $R$ over $T$ channel uses should be the same (i.e., for fair comparison with respect to the amount of information). The version of the source signal transmitted by the relay depends on the cooperative strategy, e.g., AF or DF. The channel gains $h_i,\ i \in \{\mathrm{sd}, \mathrm{sr}_1, \ldots, \mathrm{sr}_K, \mathrm{r}_1\mathrm{d}, \ldots, \mathrm{r}_K\mathrm{d}\}$, between two nodes are modeled as independent, zero-mean, circularly-symmetric complex Gaussian random variables that remain constant for the duration of one transmission block of $T$ channel uses. At each receiving node white Gaussian noise is added and noise realizations are assumed to be independent and identically distributed (i.i.d.) and $\mathcal{CN}(0, N_0)$. An average transmit power constraint of $P$ is used at the source and each relay over a transmission block, respectively, and SNR is defined as $\mathsf{SNR} = P/N_0$. We use $\epsilon$-outage capacity $\mathcal{C}_\epsilon(\mathsf{SNR})$ as performance metric, which is defined as the highest rate $R$ such that outage probability satisfies $p_{\mathrm{out}}(R, \mathsf{SNR}) := \Pr(\mathcal{C}(\mathsf{SNR}) < R) \leq \epsilon$, with $0 \leq \epsilon \leq 1$ and $\mathcal{C}(\mathsf{SNR})$ being the instantaneous channel capacity. For a given $\epsilon$, we have

$$\mathcal{C}_\epsilon(\mathsf{SNR}) := \sup\{R :\ p_{\mathrm{out}}(R, \mathsf{SNR}) \leq \epsilon\}. \tag{1}$$

We now review some important results from literature. For DF with incremental relaying in the one-relay case, $\epsilon$-outage capacity can be expressed as [2]

$$\mathcal{C}_\epsilon^{\mathrm{DF}} = \frac{1}{\mathbb{E}(N)} \log_2 \left(1 + \mathsf{SNR}\sqrt{\frac{2\sigma_{\mathrm{sd}}^2\sigma_{\mathrm{sr}}^2\sigma_{\mathrm{rd}}^2\epsilon}{2\sigma_{\mathrm{rd}}^2 + \sigma_{\mathrm{sr}}^2}}\right), \tag{2}$$

where $\mathbb{E}(N)$ denotes the average amount of transmission phases given by $\mathbb{E}(N) = 1 + \Pr(\text{"source transmission fails"})$, i.e., $\mathbb{E}(N) = 1 + \Pr(|h_{\mathrm{sd}}|^2 < (2^{2R} - 1)/\mathsf{SNR})$. For BAF with incremental relaying, $\epsilon$-outage capacity is [3]

$$\mathcal{C}_\epsilon^{\mathrm{BAF}} \approx \frac{1}{\mathbb{E}(N)} \log_2 \left(1 + \mathsf{SNR}\sqrt{\frac{2\sigma_{\mathrm{sd}}^2\sigma_{\mathrm{sr}}^2\sigma_{\mathrm{rd}}^2\epsilon}{\sigma_{\mathrm{rd}}^2 + \sigma_{\mathrm{sr}}^2}}\right). \tag{3}$$

In the above mentioned cases the feedback link has been assumed to be perfect. Therefore, the question addressed in the following is how imperfect feedback influences the $\epsilon$-outage capacity? It is evident, that imperfect feedback only influences the average amount of transmission phases $\mathbb{E}(N)$ and not the $\log$-expression. This is due to the fact that $\epsilon$-outage capacity of incremental relaying is derived by using a "baseline model", i.e., a similar network with similar relay strategy but without feedback. Feedback then comes into play by introducing a scaling factor, that depends on the successful source-to-destination transmission (for the one-relay case). $\epsilon$-outage capacity will be reduced if the average amount of transmission phases increases. For the one-relay case, this average value only depends on the source-to-destination link. To conclude, in order to investigate the influence of imperfect feedback on the $\epsilon$-outage capacity of an incremental relaying protocol, it suffices to analyze the average amount of transmission phases $\mathbb{E}(N)$. In the following, we make the useful assumption that D knows if it has been able to decode properly (i.e., there is no such thing as "D sends a positive acknowledgment, though it has not been able to decode.").

Throughout the paper we use the following notation. $P_{\mathrm{SD}}$ describes the probability that the source-to-destination transmission has been successful. Accordingly, $\bar{P}_{\mathrm{SD}}$ is the probability that the source-to-destination transmission has *not* been successful. $P_{\mathrm{R}_k\mathrm{D}}$ is the probability that D can decode after the $k$-th relay has transmitted ($k = 1, \ldots, K$). This also includes preceding transmissions. For instance, consider $P_{\mathrm{R}_1\mathrm{D}}$. This describes the probability that D can decode after combining the source's and the first relay's transmissions. Combining here depends on several aspects, e.g., the coding strategy or the cooperative protocol. For DF, D is able to decode whenever

$$|h_{\mathrm{sd}}|^2 + |h_{\mathrm{rd}}|^2 \geq \frac{2^{2R} - 1}{\mathsf{SNR}}, \tag{4}$$

where we assume that the relay has been able to decode the source transmission. For AF, D can decode if

$$|h_{\mathrm{sd}}|^2 + \frac{|h_{\mathrm{sr}}|^2|h_{\mathrm{rd}}|^2}{|h_{\mathrm{sr}}|^2 + |h_{\mathrm{rd}}|^2 + 1/\mathsf{SNR}} \geq \frac{2^{2R} - 1}{\mathsf{SNR}}. \tag{5}$$

Consequently, $\bar{P}_{\mathrm{R}_k\mathrm{D}}$ is the probability that D *cannot* decode after the $k$-th relay has transmitted. A positive acknowledgment from D is denoted by `ACK` and a negative acknowledgment by `NACK`. With $(\mathbf{AB})_{l,m}$ we denote the element of the $l$-th row and the $m$-th column of the matrix product $\mathbf{AB}$. This means for

$$\mathbf{A} = \begin{bmatrix} a_0 & a_1 \\ a_2 & a_3 \end{bmatrix}, \quad \mathbf{B} = \begin{bmatrix} b_0 & b_1 \\ b_2 & b_3 \end{bmatrix},$$

we have

$$(\mathbf{AB})_{2,1} = a_2 b_0 + a_3 b_2.$$

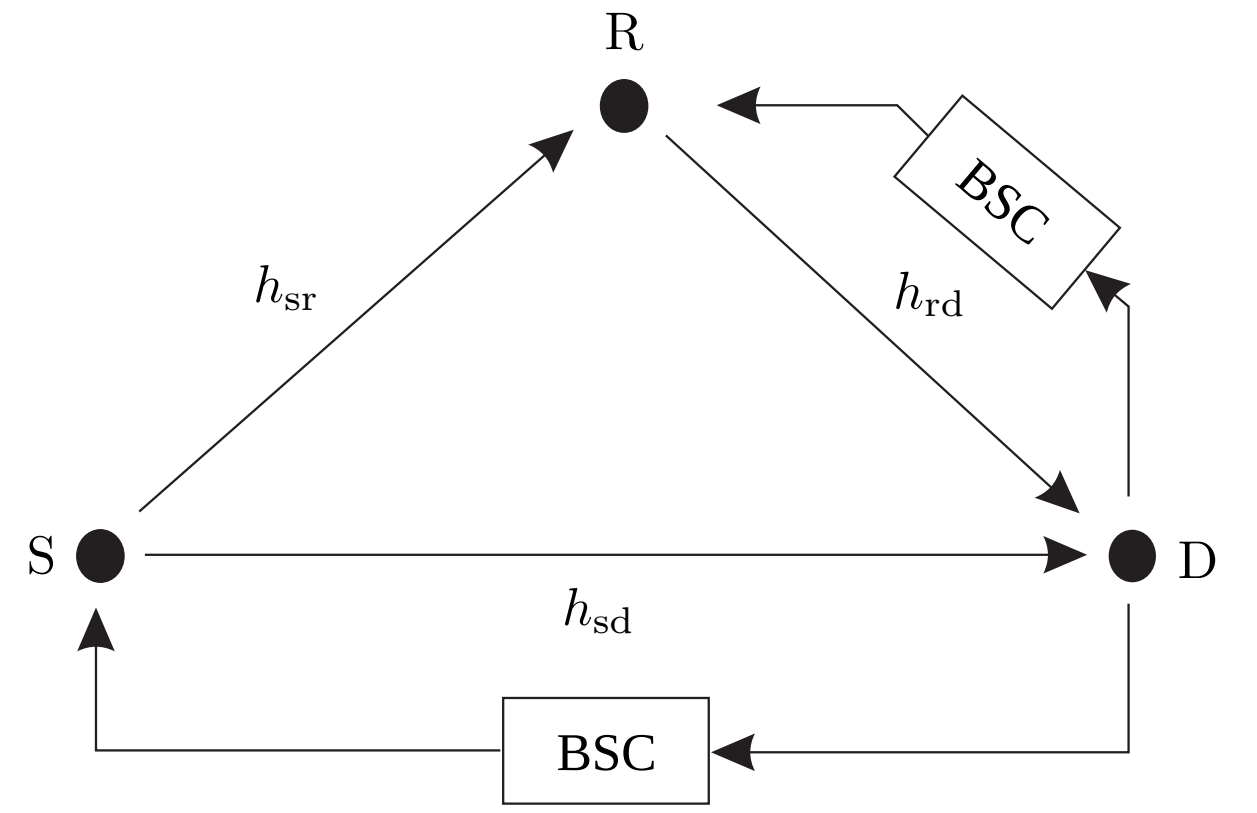

Fig. 1. Network for incremental relaying with imperfect feedback modeled as binary symmetric channel (BSC).

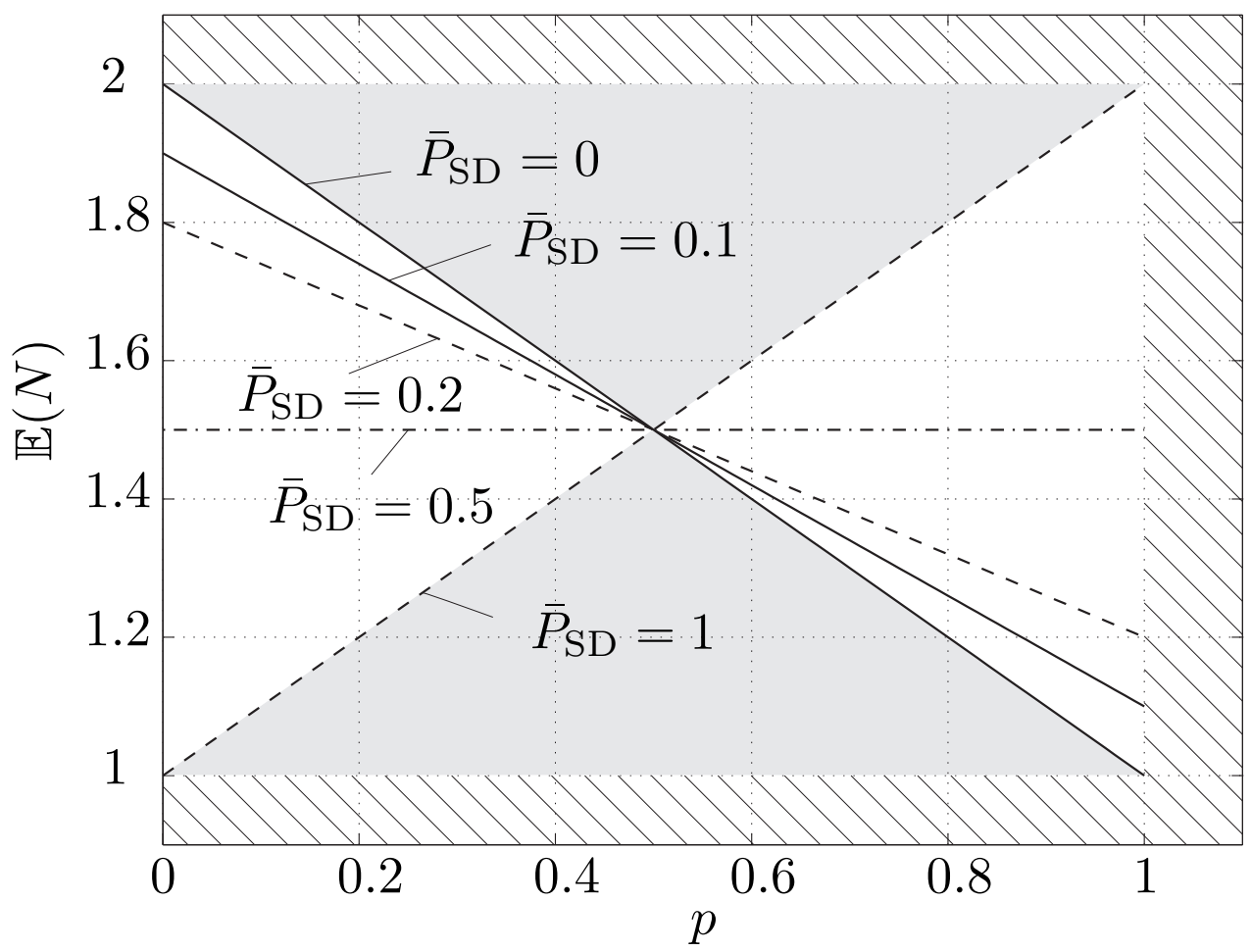

Fig. 2. Average amount of transmission phases $\mathbb{E}(N)$ versus reliability of feedback link $p$ for the one-relay case.

The Hadamard product of two matrices $\mathbf{A}$ and $\mathbf{B}$ is given by

$$\mathbf{A} \circ \mathbf{B} = \left[ \begin{array}{cc} a_0 b_0 & a_1 b_1 \\ a_2 b_2 & a_3 b_3 \end{array} \right].$$

## III. Imperfect Feedback

Since we deal with a one-bit feedback, it is reasonable to model the feedback links as binary symmetric channel (BSC) defined as

$$\begin{aligned} p &:= \Pr(\texttt{ACK}|\texttt{ACK}) = \Pr(\texttt{NACK}|\texttt{NACK}) \\ 1-p &:= \Pr(\texttt{NACK}|\texttt{ACK}) = \Pr(\texttt{ACK}|\texttt{NACK}). \end{aligned}$$

The network model for the one-relay case is illustrated in Fig. 1. For the sake of analysis, we assume that source and relays receive the feedback with the same degree of reliability, i.e., the feedback channel from D to all transmitting terminals (source and relays) is the same and, hence, modeled by the same BSC with parameter $p$.

### *A. One-Relay Case*

We first consider the one-relay case. There are two constellations that lead to one transmission phase. Either the source transmission has been successful and ACK has been received correctly or the source transmission has not been successful and NACK has been received incorrectly. Apart from that, we get two transmission phases for the following cases. Source transmission has been successful and ACK has been received incorrectly or source transmission has not been successful and NACK has been received correctly. To sum up, this can be expressed as

$$\begin{aligned} \mathbb{E}(N) &= P_{\rm SD} p + \bar{P}_{\rm SD}(1-p) + 2P_{\rm SD}(1-p) + 2\bar{P}_{\rm SD} p \\ &= (2\bar{P}_{\rm SD} - 1)p + 2 - \bar{P}_{\rm SD}, \end{aligned}$$

where we used $P_{\rm SD} = 1 - \bar{P}_{\rm SD}$ in the second line. This clearly represents a linear equation of $\mathbb{E}(N)$ in $p$ depending on the parameter $\bar{P}_{\rm SD}$. All curves are in a rectangular box bounded by $\mathbb{E}(N) = 1$, $\mathbb{E}(N) = 2$, $p = 0$, and $p = 1$. The array of curves for different values of $\bar{P}_{\rm SD}$ is illustrated in Fig. 2. Clearly, the extreme values of $\bar{P}_{\rm SD}$ are $\bar{P}_{\rm SD} = 0$ and $\bar{P}_{\rm SD} = 1$, which bound the regions in which we cannot find any curves (gray area). It can be seen that depending on the parameter $\bar{P}_{\rm SD}$, $\mathbb{E}(N)$ increases or decreases in $p$. In order to investigate this behavior, we calculate the derivation of $\mathbb{E}(N)$ with respect to $p$ and have:

$$\frac{\mathrm{d}\mathbb{E}(N)}{\mathrm{d}p} = 2\bar{P}_{\rm SD} - 1 \left\{ \begin{array}{l} < 0 \text{ for } \bar{P}_{\rm SD} \in [0; 0.5) \\ > 0 \text{ for } \bar{P}_{\rm SD} \in (0.5; 1] \end{array} \right.$$

For $p = 1$, we have "perfect" feedback and the average amount of transmission phases is $\mathbb{E}(N) = 1 + \bar{P}_{\rm SD}$. This is the case investigated in [2] for DF and in [3] for BAF. For $p = 0$, where each observation of feedback is wrong with probability 1, we have $\mathbb{E}(N) = 2 - \bar{P}_{\rm SD}$. Moreover, an interesting fact is that all curves intersect at $(p = 0.5; \mathbb{E}(N) = 1.5)$. Why is this the case? For $p = 0.5$, observation of feedback is worthless. Therefore, from a long-term perspective, it is best if the relay transmits in block $i$, remains silent in block $i+1$, transmits in block $i+2$ and so on. This strategy eventually leads to $\mathbb{E}(N) = 1.5$ independent of $\bar{P}_{\rm SD}$. Or - in other words - the relay scrambles in each block if it should transmit or not.

We summarize our results in a few words.

- If $\bar{P}_{\rm SD} < 0.5$, then $\mathbb{E}(N)$ decreases with increasing $p$. Therefore, if the S-to-D link is reliable (i.e., $\bar{P}_{\rm SD} \to 0$), the average amount of transmission phases $\mathbb{E}(N)$ decreases, when the feedback channel gets more and more reliable (i.e., $p \to 1$).
- If the S-to-D link is *not* reliable (i.e., $\bar{P}_{\rm SD} \to 1$), the average amount of transmission phases $\mathbb{E}(N)$ increases, when the feedback channel gets more and more reliable (i.e., $p \to 1$). This is also intuitively clear. If S-to-D fails pretty often and the relay receives information from the destination about success or failure of S-to-D transmission correctly, the relay has to aid communications more often, hence, $\mathbb{E}(N)$ increases with increasing $p$.

- Consider the case $\bar{P}_{\mathrm{SD}} = 0.5$ (dash-dotted line in Fig. 2). The relay should transmit in every second block (from a long-term perspective, since every second source transmission fails). Therefore, $\mathbb{E}(N)$ becomes 1.5.

The average amount of transmission phases can also be expressed in matrix notation:

$$\mathbb{E}(N) = \underbrace{[1,2]}_{\mathbf{K}_2} \underbrace{\begin{bmatrix} p & 1-p \\ 1-p & p \end{bmatrix}}_{\mathbf{P}} \underbrace{\begin{bmatrix} P_{\mathrm{SD}} \\ \bar{P}_{\mathrm{SD}} \end{bmatrix}}_{\mathbf{S}} = \mathbf{K}_2\mathbf{P}\mathbf{S}$$

The vector $\mathbf{K}_2$ describes the possible amount of transmission phases. The matrix $\mathbf{P}$ describes the feedback link modeled as BSC and $\mathbf{S} = [P_{\mathrm{SD}}\ \bar{P}_{\mathrm{SD}}]^T$ is a vector denoting successful or failed S-to-D transmission.

### *B. Generalization*

Due to the structure of how $\mathbb{E}(N)$ is calculated, we are able to derive a binary tree-based construction rule for generalized networks with an arbitrary number of relay nodes. For that, consider Fig. 3. The left figure (subfigure a)) shows the one-relay case and the right figure (subfigure b)) illustrates the two-relay case. We immediately see that the number of "levels" corresponds to the number of relays, i.e., for the one-relay case we have only one level and for the two-relay case the number of levels becomes 2. Furthermore, there are two kinds of blocks. A "positive" block that deals with successful transmission (e.g., $P_{\mathrm{SD}}$ and $P_{\mathrm{R_1D}}$) and the possibility of successful or failed positive acknowledgment (ACK). And a "negative" block that considers failed transmission (e.g., $\bar{P}_{\mathrm{SD}}$ and $\bar{P}_{\mathrm{R_1D}}$) and the possibility of successful or failed negative acknowledgment (NACK). We talk of a "path", when we consider the multiplication of a decoding probability with the corresponding ACK or NACK. In each level, both kinds of blocks appear. In order to derive $\mathbb{E}(N)$, the following construction rule can be applied that gives the different summands during calculation.

*1) Positive block:*

- If it ends with $p = \Pr(\texttt{ACK}|\texttt{ACK})$, then this path is terminated and multiplied by the level number. For the one-relay case, this is the path $P_{\mathrm{SD}}p$. For the two-relay case, these are the paths $P_{\mathrm{SD}}p$ which is multiplied by 1, as well as $P_{\mathrm{SD}}(1-p)P_{\mathrm{R_1D}}p$ and $\bar{P}_{\mathrm{SD}}pP_{\mathrm{R_1D}}p$ which are multiplied by 2.
- If it ends with $1-p = \Pr(\texttt{NACK}|\texttt{ACK})$, a new level is added, i.e., a new positive and a new negative block are added. The construction continues until the highest level has been reached. (The highest level corresponds to the number of relays in the network.) Then, the last path is multiplied by a factor that is equal to the highest level number plus 1. For the one-relay case, this is the path $P_{\mathrm{SD}}(1-p)$ which is multiplied by 2. For the two-relay case, these are the paths $P_{\mathrm{SD}}(1-p)P_{\mathrm{R_1D}}(1-p)$ and $\bar{P}_{\mathrm{SD}}pP_{\mathrm{R_1D}}(1-p)$ which are multiplied by 3.

*2) Negative block:*

- If it ends with $1-p = \Pr(\texttt{ACK}|\texttt{NACK})$, then this path is terminated and multiplied by the level number. For the one-relay case, this is the path $\bar{P}_{\mathrm{SD}}(1-p)$ which is multiplied by 1. For the two-relay case, these are the paths $P_{\mathrm{SD}}(1-p)\bar{P}_{\mathrm{R_1D}}(1-p)$ and $\bar{P}_{\mathrm{SD}}p\bar{P}_{\mathrm{R_1D}}(1-p)$ which are multiplied by 2.
- If it ends with $p = \Pr(\texttt{NACK}|\texttt{NACK})$, a new level is added, i.e., a new positive and a new negative block are added. The construction continues until the highest level has been reached. Then, the last path is multiplied by a factor that is equal to the highest level number plus 1. For the one-relay case, this is the path $\bar{P}_{\mathrm{SD}}p$ which is multiplied by a factor 2. For the two-relay case, these are the paths $P_{\mathrm{SD}}(1-p)\bar{P}_{\mathrm{R_1D}}p$ and $\bar{P}_{\mathrm{SD}}p\bar{P}_{\mathrm{R_1D}}p$ which are multiplied by 3.

We see that with this rather simple construction rule, we are able to describe the average amount of transmission phases for networks with an arbitrary number of relay nodes. For $p = 0.5$, the story gets even more interesting. For the case of one relay, we have $\mathbb{E}(N) = 1.5$. For two relays, we get $\mathbb{E}(N) = 1.75$. It can easily be verified that due to the binary tree-based construction rule explained before, the limit for $K \to \infty$, where $K$ is the number of relay nodes, tends to 2. This can directly be seen if we consider the geometric series $\sum_{k=0}^{\infty} \frac{1}{2^k} = 2$. To sum up, if the feedback link is unreliable, i.e., $p = 0.5$, and we have a network with a lot of relays (i.e., $K$ large), the transmission strategy of the relays should be as follows. Each source message is retransmitted by one and only one relay, which clearly leads to $\mathbb{E}(N) = 2$. This is in line with results presented in [8], [9]. There, an opportunistic relay protocol is proposed, where only one relay out of $K$ is used for cooperation.

With respect to Fig. 3, we are able to express $\mathbb{E}(N)$ for networks with an arbitrary number of $K$ relays in matrix notation. The key is to exploit the binary construction rule and to keep in mind that per level there are $2^k$ paths that are terminated and multiplied by the level number, i.e., they do not have to be considered for further calculations anymore. The result is shown in Fig. 4. We see that the resulting matrix is multiplied by a $1 \times (K+1)$ vector $\mathbf{K}_{K+1}$.

## IV. Conclusions and Outlook

In this paper we dealt with the effect of imperfect feedback on the $\epsilon$-outage capacity of incremental relaying in the low signal-to-noise ratio (SNR) regime by modeling the feedback links as binary symmetric channel (BSC). It has been shown that imperfect feedback influences the pre-log factor of capacity expressions in a way that it reduces the $\epsilon$-outage capacity. In addition to that, the effect of different degrees of feedback reliability on the system performance have been investigated. It has been shown that for large networks the average amount of transmission phases in order to send one source message tends to 2 if the observation of feedback is worthless. This result leads to a transmission strategy, where only one relay out of $K$ is retransmitting the source message. Furthermore, due to a binary tree-based construction rule for the calculation of the average amount of transmission phases, a compact matrix notation for networks with an arbitrary number of relay nodes could be given. In the present considerations, we assumed that each feedback link is the same. However, it would be more realistic to model each feedback link as a different BSC. This extension is straightforward to the presented work.

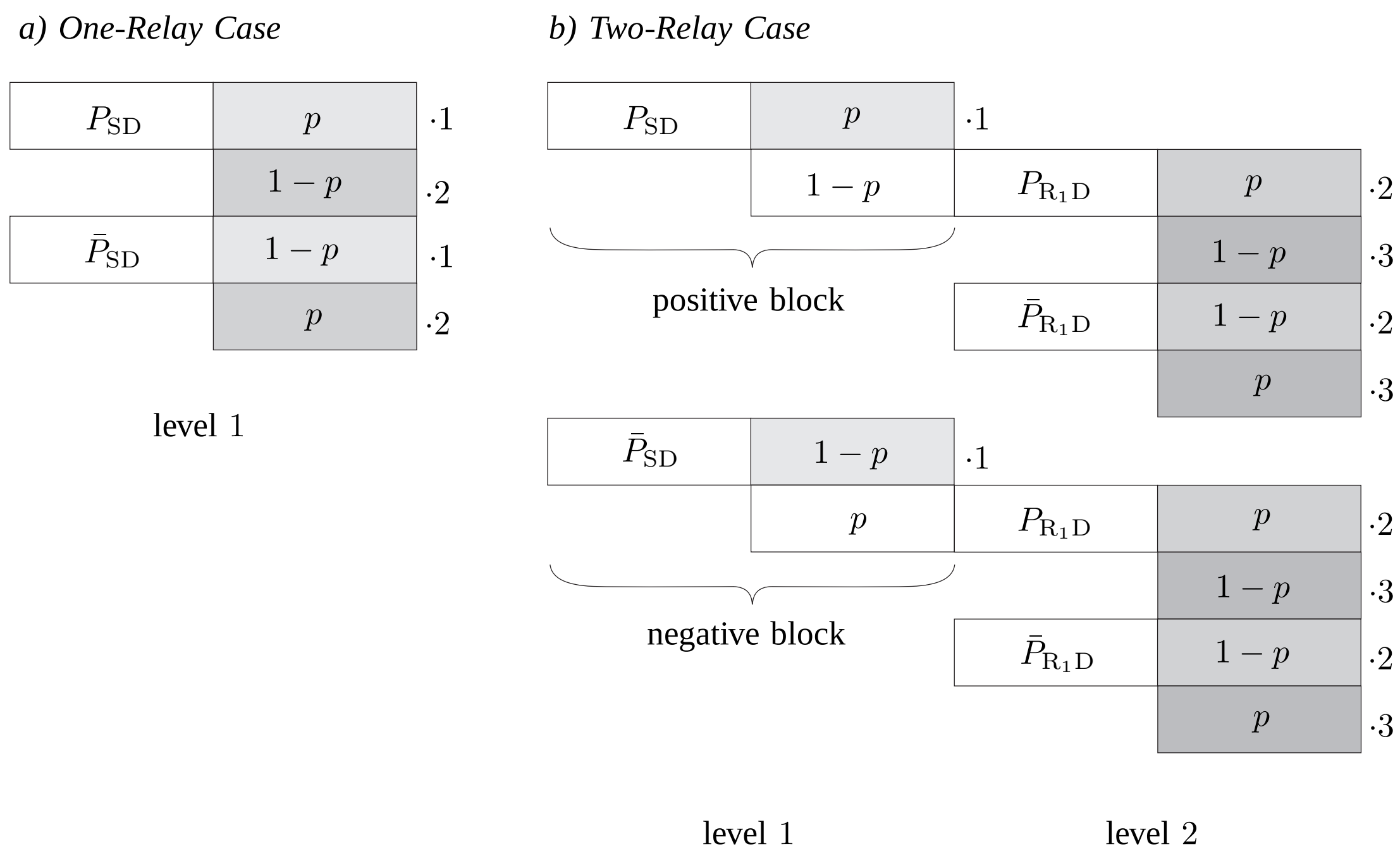

Fig. 3. Binary tree-based construction in order to calculate the average amount of transmission phases $\mathbb{E}(N)$. a) One-relay case and b) two-relay case. Number of levels corresponds to the number of relays in the network.

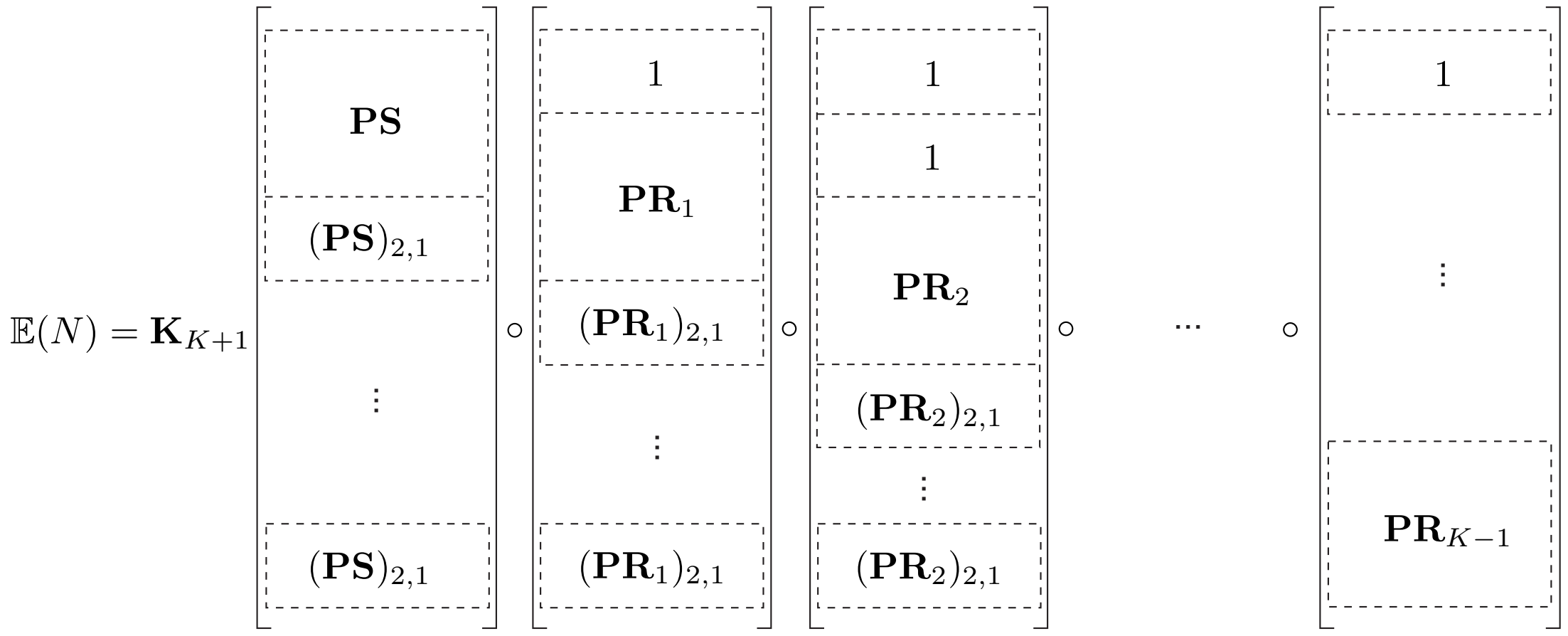

Fig. 4. Matrix notation of $\mathbb{E}(N)$ for the case of $K$ relays.